\documentclass[prl,twocolumn,superscriptaddress,showpacs]{revtex4}

\usepackage{amsmath}
\usepackage{graphicx}
\usepackage{amssymb}
\usepackage{amsthm, amscd}
\usepackage{amsfonts}
\usepackage{times}
\usepackage{pstricks}
\usepackage{wasysym}

\begin{document}

\title{Bulk-edge coupling in the non-abelian $\nu=5/2$ quantum Hall interferometer}

\author{ B. Rosenow}\thanks{On leave from the Institut f\"ur Theoretische Physik, Universit\"at zu
K\"oln, D-50923, Germany}
\affiliation{ Physics Department, Harvard University, Cambridge
02138, MA}
\author{ B. I. Halperin}
\affiliation{ Physics Department, Harvard University, Cambridge
02138, MA}
\author{S. H. Simon}
\affiliation{Lucent-Alcatel Bell Laboratories, Murray Hill NJ, 07974}
\author{Ady Stern}
\affiliation{ Department of Condensed Matter Physics, Weizmann
Institute of Science, Rehovot 76100, Israel}

\date{July 30, 2007}

\begin{abstract}
Recent schemes for experimentally probing non-abelian statistics in
the quantum Hall effect are based on geometries where current-carrying
quasiparticles flow along edges that encircle bulk quasiparticles,
which are localized. Here we consider one such scheme, the Fabry-Perot
interferometer, and analyze how its interference patterns are affected
by a coupling that allows tunneling of neutral Majorana fermions
between the bulk and edge.  While at weak coupling this tunneling
degrades the interference signal, we find that at strong coupling, the
bulk quasiparticle becomes essentially absorbed by the edge and the
intereference signal is fully restored.
\end{abstract}

\pacs{73.43.Cd, 73.43.Jn}

\maketitle

Recently, interference experiments were proposed as a way to examine
the non-abelian nature of quasiparticles in the $\nu=5/2$ quantum Hall
state\cite{review,stern06,bonderson06,dassarma05}. The most dramatic
signature of non-abelian statistics is expected to be seen in the
interference of back-scattering amplitudes from two constrictions in a
long Hall bar.  (See inset in Fig.~1.) The two constrictions enclose a
``cell", whose area may be varied by means of a side-gate.  The bulk
is assumed to host a number $N_{qp}$ of localized quasiparticles, that
do not take part in electronic transport, and have no tunnel coupling
to the edge.  In the limit of weak back-scattering, when $N_{qp}$ is
even the two back-scattering amplitudes interfere coherently, while
when $N_{qp}$ is odd they are incoherent, and thus do not
interfere. In the former case, the back-scattered current oscillates
with the area of the cell, while in the latter case it does not.  This
difference reflects the non-abelian nature of the quasiparticles.

The theoretical analysis makes a sharp distinction between bulk and
edge.  In a real system, however, some degree of coupling between the
edge and quasiparticles localized in the bulk is unavoidable.  Since
both the $5/2$ edge and the quasiparticles consist of both neutral
Majorana fermionic and charged bosonic degrees of freedom, several
types of edge to bulk coupling are possible. We expect that at low
energies, tunneling that involves a charge will generally be
suppressed due to the Coulomb energy. Thus in this work we will focus
on tunneling of the neutral Majorana mode from the bulk to the edge,
and on the resulting effect on the interference.

The system we consider \cite{chamon97,fradkin98,stern06,bonderson06}
is a Hall bar lying parallel to the $x$--axis (See Fig.~1).  Two
constrictions are located at $x=-b$ and $x=b$. We focus on a simple
case where there are two quasiparticles, $N_{qp}=2$, localized at
$x=0$, between the two constrictions, with one of the quasiparticles
coupled to the upper edge and the other coupled to the lower edge.
Ref. \cite{Wen} considers the case of $N_{qp}=1$ in the weak tunneling
limit.

When the two localized quasiparticles are decoupled from the edge they
form a two level system, and the ground state is doubly
degenerate. The interference patterns that are seen in the two
respective ground states are mutually shifted by a phase $\pi$.  Then,
at temperature $T=0$ the magnitude of the interference term depends on
the ratio of two time scales, one determined by the voltage
$t_V=\hbar/e^*V$, and the other being the time associated with motion
between the two constrictions $t_b = 2b/v$ where $v$ is a
characteristic edge mode velocity.  When analyzing the effect of
bulk-edge coupling we will focus on the case of low voltage, $t_V\gg
t_b$, where the interference is most clearly seen.

We start with a qualitative description of our results.  In the
absence of edge-bulk coupling the system cannot switch from one ground
state to another. Thus if it is prepared in one ground state,
repetitive measurements of the interference would show the same
interference pattern.  However, when the interference term is averaged
over the two possible ground states, e.g., by measuring the
interference with a random choice of the initial ground state, the
average is zero.  When the coupling of the bulk two-level system to
the edge is turned on, the average value of the interference term
becomes non-zero, and the correlation function between consecutive
measurements is strongly modified.  Denoting the coupling strengths
between the localized Majorana particles and their respective edges by
$\lambda_u$ and $\lambda_d$, [defined precisely in
Eq.~(\ref{bulk-edge-tunneling}) below], we obtain corresponding time
scales $t_{\lambda u(d)} = (
\pi \lambda^{2}_{u(d)}/2 v_m)^{-1}$, where $v_m$ is the velocity of
the Majorana modes on the edges.  In the limit of weak coupling,
where $t_{\lambda u(d)} \gg t_V$, we may use a perturbation
analysis, and we find that the average value of the interference is
proportional to $(t_V)^{1/2}(t_V/t_\lambda) \log^2[t_\lambda/t_V]$,
where we have assumed that $t_{\lambda u}$ and $t_{\lambda d}$ are
comparable in magnitude, and $t_{\lambda }$ is their geometric mean.
As the coupling is increased, or as the voltage is
lowered, the perturbative analysis breaks down. We then carry out a
numerical analysis, which suggests that in the limit
$t_V/t_\lambda\rightarrow \infty$ the full magnitude of the
interference term is retrieved. In effect, the two bulk
quasiparticles become then a part of the edge, and $N_{qp}$ reduces
from two to zero. (We find a similar effect for a single
quasiparticle strongly coupled to an edge.) In contrast to the
build-up of the average interference term as the coupling gets
stronger, the fluctuating part of the interference pattern is
weakened by the coupling, and its  characteristic correlation time
becomes $t_\lambda$, which decreases with increased coupling.

For the derivation of these results we will follow several steps.
After the introduction of the relevant Lagrangians, we derive the
operator that describes quasiparticle tunneling across the two
constrictions in the presence of the two localized bulk
quasiparticles. We find it useful to represent this operator in two
forms, a local form using the $\sigma$ operator of the Ising Conformal
Field Theory (CFT) that describes the $\nu=5/2$ edge, and a non-local
form in terms of the Majorana fermions that propagate along the
edge. Within the non-local form, we show that the tunneling operator
is proportional to a ``parity operator" that measures the parity of
the number of electrons encircled by a back-scattered quasiparticle as
it moves from $x=-\infty$ along one edge, through the back-scattering
at the constriction, back to $x=-\infty$ along the other edge.  Next,
we perturbatively analyze the weak coupling limit, and finally we
numerically analyze the strong coupling limit.

In the absence of any coupling to bulk quasiparticles the upper ($u$)
and lower ($d$) edges of the $\nu=5/2$ state are described by two
charged boson fields $\phi_u(x),\phi_d(x)$ and a neutral Majorana
fermion field. The Lagrangian for the boson field on each edge is that
of a chiral Luttinger liquid, characterized by a velocity $\pm v_c$.
The Lagrangian density for the Majorana fermion field is ${\cal
L}_m^r={1 \over 4 \pi} \psi^r(x)[\partial_t-v_m^r\partial_x]\psi^r(x)$
with $r$ taking the values $u$ and $d$ for the upper and lower
edges. For simplicity we set the velocities of the Majorana edge modes
to be equal and opposite $v_m^u=-v_m^d=v_m$. Furthermore, we set
$v_m=1$ when no confusion results.  The Majorana Lagrangian can also
be thought of as the Lagrangian of an Ising CFT \cite{YellowBook}.

Each of the two localized bulk quasiparticles carries a zero mode,
described by a localized Majorana operator. We denote the two bulk
Majorana operators by $\gamma_u,\gamma_d$, with the subscript
indicating the edge to which the quasiparticle couples. The
two-dimensional Hilbert space created by the two Majorana modes is
spanned by the two eigenvectors of the operator $i\gamma_u\gamma_d$.

To examine the effects of bulk-edge coupling we couple $\gamma_u$ to
the upper edge and $\gamma_d$ to the lower edge, both at $x=0$. The
Lagrangian density for this coupling is
\begin{equation}
{\cal L}_{b-e}=
i\left[\lambda_u\psi^u(x)\gamma_u+\lambda_d\psi^d(x)\gamma_d\right
]\delta(x) \ \ .
 \label{bulk-edge-tunneling}
\end{equation}
The Lagrangian ${\cal L}_m^u+{\cal L}_m^d+{\cal L}_{b-e}$ introduces
the time scales $t_{\lambda u(d)}$ defined above.  The bulk-edge
coupling mixes the states with eigenvalues $\pm 1$ of
$i\gamma_u\gamma_d$. Roughly speaking, $t_\lambda$ is the time in
which a state with a particular value of $i\gamma_u\gamma_d$ decays to
a mixture of the two eigenvalues.

The operator that tunnels a quasiparticle across a constriction may be
expressed in a local form through the $\sigma_u, \sigma_d$ operators
of the Ising CFT that describes the $\nu=5/2$ upper and lower edges
\cite{SBPRB}. The tunnelling operator is $H_{tun}\equiv {\hat T}
+{\hat T}^+ $, where
\begin{equation}
{\hat T}=e^{i e^* V t} \left[ \eta_L  {\cal C}_L{\cal N}_L +\eta_R {\cal
C}_R{\cal N}_R i \gamma_u\gamma_d \right]
\label{tunneling-schematic}
\end{equation}
transfers a quasiparticle from the lower to the upper edge through the
left $(L)$ and right $(R)$ constrictions respectively, and its
hermitian conjugate ${\hat T}^+$ similarly transfers a quasiparticle
from the upper to the lower edge.  Here, $V$ is the voltage difference
between the two edges, $e^* = e/4$ is the quaisparticle
charge. Correspondingly, the current operator is given by $J =
\frac{e^*}{i}(T - T^+)$.  The operators ${\cal C}_{L(R)}\equiv
e^{i\left (\phi_u(\mp b)-\phi_d(\mp b)\right )/\sqrt{8}}$ are the
charge part of the tunneling operator, operating on the charge
mode. The Aharonov-Bohm phase is absorbed into the relative phase
between the tunneling coefficients $\eta_{L,R}$. The neutral parts of
the tunneling operators are ${\cal N}_L\equiv
\sigma_u(-b)\sigma_d(-b)$ and ${\cal N}_R\equiv
\sigma_u(b)\sigma_d(b)$.  For the present purpose, the $\sigma$
operators are defined through their operation on the Majorana
fermion fields as \cite{SBPRB}
\begin{equation}
\sigma_r(x_0)\psi_{r}(y)\sigma_r(x_0)= -{\rm sgn}(x_0-y)\psi_{r}(y)
\label{sigmadef}
\end{equation}
with $r=u,d$.  The factor of $\gamma_u \gamma_d$ in the second term of
Eq.~(\ref{tunneling-schematic}) is included to account for the
wrapping of a tunneling quasiparticle at position $x=b$ around the two
localized quasiparticles. This factor is responsible for the $\pi$
phase shift between the interference patterns corresponding to the two
eigenvectors of $\gamma_u\gamma_d$.

The neutral mode part of the tunneling operators may also be expressed
in a non-local form through the Majorana fermions along the two edges
in a way which we find to be both illuminating and useful. This
approach is based on the description of the $\nu=5/2$ state as a
$p$--wave superconductor of composite fermions \cite{review}.  Within
this description the bulk is a superconductor, with the localized
quasiparticles being vortices in that superconductor.  A tunneling of
a quasiparticle from one edge to another at position $x_0$ involves a
tunneling of a vortex, and that introduces a twist into the phase of
the order parameter: for all points in the region $x<x_0$, the phase
is shifted by $2\pi$, while for all points in the region $x>x_0$ the
phase is unaffected by the vortex motion (up to an unimportant global
gauge redefinition). To implement this shift of the phase, we
recognize that the phase field is canonically conjugate to the
Cooper-pair density field, which at zero temperature is just half the
electron density field. The operator that implements the required
shift in the phase is then
\begin{equation}
P(-\infty,x_0)=e^{i \pi \int_{x\le x_0} d{\bf r} \,
\rho({\bf r})} \, .  \label{parityoperator}
\end{equation}
Since the operator $\int_{x\le x_0} d{\bf r} \, \rho({\bf r})$ has
only integer eigenvalues, the operator $ P(-\infty,x_0)$ is nothing
but a {\it Parity Operator} which measures the parity of the number of
electrons to the left \cite{endnote2} of $x_0$.
Eq.~(\ref{tunneling-schematic}) can thus be rewritten as $\hat T =
e^{i e^* V t} \left[\eta_L{\cal C}_L P(-\infty,-b) + \eta_R {\cal C}_R
P(-\infty,b)\right]$ as we shall see below.

Since the bulk of the system is gapped, and since all particles in the
superconducting ground state are paired, the parity operator only has
contributions from localized neutral modes and from the neutral mode
along the edge. The operator $i\gamma_u\gamma_d$ in the second term of
Eq.~(\ref{tunneling-schematic}) precisely counts the parity of the
number of fermions in the localized bulk quasiparticles to the left of
$x=b$.  Counting the fermions along the edge is a bit more complicated
but is achieved by constructing a complex Fermi field
$\psi_e(x)=\psi_u(x)-i\psi_d(x)$ and
$\psi_e^\dagger(x)=\psi_u(x)+i\psi_d(x)$, such that the edge
contribution to the parity operator is
\begin{equation}
P_{edge}(-\infty,x_0)=e^{i\pi\int_{-\infty}^{x_0} dx \,
\psi_e^\dagger(x)\psi_e(x)}. \label{edgeparity}
\end{equation}
It is easy to see that Eq.~(\ref{sigmadef}) holds when the operators
$\sigma_r(x_0)$ are replaced by $P_{edge}$.  The eigenvalues of the
latter are $\pm 1$, since the eigenvalues of $\int^{x_0}dx
\psi_e^\dagger(x) \psi_e(x)$ are integers. The application of either
$\psi_d(y)$ or $\psi_u(y)$ on an eigenstate of $\int^{x_0}dx
\psi_e^\dagger(x) \psi_e(x)$ changes the eigenvalue by $\pm
\theta(x_0-y)$, and hence (\ref{sigmadef}).  Altogether, then, we have
${\cal N}_{L(R)}=P_{edge}(-\infty,\mp b) $.

To calculate the current-voltage characteristics in the weak
back-scattering limit, we use standard \cite{chamon97} perturbation
theory in the tunneling strength to yield $ I = -i \int_{-\infty}^0
\!\! dt \, \, \langle [J(0), H_{tun}(t)] \rangle $.  With some
algebra, the interference term that results is
\begin{eqnarray}
I_{int} &=&{\rm Re}\frac{2e^*\eta_R \eta _L^*}{\hbar}\int_{-\infty}^\infty dt \, e^{-ie^*Vt}\nonumber\\
&\times &\left \langle \left[{\cal C}_L^+(t){\cal N}_L(t), {\cal
C}_R(0){\cal N}_R(0)\gamma_u(0)\gamma_d(0)\right ] \right\rangle \  \ .
 \label{interference}
\end{eqnarray}
For $- e^\star V >0$, only the first term of the commutator, with
$t$-dependent operators to the left, will contribute to the integral.
The correlator of the charged operators (the ${\cal C}$'s) and that of
the neutral operators (the ${\cal N}$'s and $\gamma$'s) factorize. The
correlator of the charged operator is $\langle {\cal C}^+_L(t){\cal
C}_R(0)\rangle = [\delta+i(v_ct - 2 b)]^{-1/8}[\delta+i(v_ct + 2
b)]^{-1/8}$, where $\delta$ is a short-distance cutoff.  The bulk-edge
coupling affects only the neutral correlator, to be denoted by ${\cal
I_N}(t) = \langle {\cal N}_L(t) {\cal N}_R(0)\gamma_u(0)\gamma_d(0)
\rangle$ which is just the parity-parity correlator $\langle
P(-\infty,-b;t) P(-\infty,b; t=0)
\rangle$. In the absence of
edge-bulk coupling, this correlator
breaks into a product 
$\langle{\cal N}_L(t){\cal
N}_R(0)\rangle_0 \,    \langle\gamma_u\gamma_d\rangle_0 $. We denote ${\cal I_N}^{(0)} \equiv \langle{\cal N}_L(t){\cal
N}_R(0)\rangle_0 $
which has the value
 $[\delta+i(t - 2 b)]^{-1/8}[\delta+i(t + 2 b)]^{-1/8}$
\cite{SBPRB,YellowBook}. The correlator
$\langle\gamma_u\gamma_d\rangle_0 $ is $\pm i$, depending which
ground state is considered.  For either ground state, integration of
these two expressions in Eq.~(\ref{interference}) leads to an
interference term of the same visibility as in the absence of any
bulk quasiparticles \cite{ardonne07}. Since we are interested in the
effect of the bulk-edge coupling on the visibility of the
interference, we find it useful to define a {\it reduction factor}
$R(t)\equiv {\cal I_N}(t)/{\cal I_N}^{(0)}(t)$.

We now turn to analyze the reduction factor in various regimes of
bulk-edge coupling. Generally, the two edge theories ($u,d$) factorize
and we can write the correlator ${\cal I_N}(t) = {\cal I}_u(t) {\cal
I}_d(t)$.  In the limit of weak coupling, we may use perturbation
theory. We expand the time evolution operator to lowest order in
$\lambda$. The perturbed correlators can be written as correlators in
an unperturbed theory
\begin{eqnarray}
\label{eq:Ivalue}
    {\cal I}_u(t) = & & \\
    \lambda_u \!\! \int \! dt'   & & \!\!\!\! \langle  {\cal T} \sigma_u(-b,t)
 \sigma_u(b,0)  \psi_u(0,t') \rangle_0 \,\, \langle   {\cal T} \gamma_u(0)
\gamma_u(t')\rangle_0 \nonumber
\end{eqnarray}
where the time integration contour starts at $-\infty$ goes up to $t$
across the real axis then back to $-\infty$ and $\cal T$ represents
the appropriate (Keldysh) time ordering of operators.

The correlators of the type $\langle\sigma\sigma\psi\rangle_0$ are
well known from conformal field theory \cite{YellowBook}: $\langle
\sigma(z_1)\sigma(z_2)\psi(z_3)\rangle_0 =
\frac{1}{\sqrt{2}}(z_{12})^{3/8}\left (z_{23}z_{13}\right )^{-1/2}$
where $z_{ij}\equiv z_i-z_j$ and $z = x + i \tau$ in imaginary time,
which then needs to be continued back to real time.  Substituting this
correlator in Eq.~(\ref{eq:Ivalue}) and noting that at the unperturbed
level $\langle\gamma_u(0)\gamma_u(t)\rangle_0 =1$ we find the time
integral to be logarithmically divergent.  However, when the
correlator $\langle\gamma_u(0)\gamma_u(t)\rangle$ is itself calculated
in perturbation theory, it is found to decay at a time scale of order
$t_\lambda$. Thus this correlator provides a natural cutoff for the
time integration. Evaluating the integrals with the cutoff yields (in
the limit of small $t_b$) that the leading contribution of the upper
edge to the parity correlator (which is independent of the details of
the cutoff) is
\begin{equation}
 {\cal I}_u (t) = (\delta + it)^{3/8}\{  \lambda_u \sqrt{2} [-i \log (|t|/t_{\lambda_u})  - \pi \mbox{sgn}
 t] \} \label{eq:Iu}.
\end{equation}

When we consider coupling of impurities to both edges, we obtain a
similar expression for ${\cal I}_d$.  The reduction factor defined
above is then
\begin{equation}
\label{eq:CFTresult} R(t) =  2 \lambda_u \lambda_d t \log
(|t|/t_{\lambda_u}) \log (|t|/t_{\lambda_d}) + \ldots
\end{equation}
Including the contributions from the charge modes and ${\cal I}_{\cal
N}^0$ in Eq.~(\ref{interference}), results in an interference current
proportional to $\lambda_u
\lambda_d V^{-3/2} \log (t_{\lambda_u}) \log (t_{\lambda_d})$. Interestingly, in the case where there is only a single
bulk quasiparticle, coupled to just one edge, the corresponding
logarithmic factor disappears from the interference current, which, in
the weak coupling limit, is proportional to $\lambda V^{-1}$, as was
shown in Ref.~\cite{Wen,endnote3}.

\begin{figure}[h]
\includegraphics[width=0.95\linewidth]{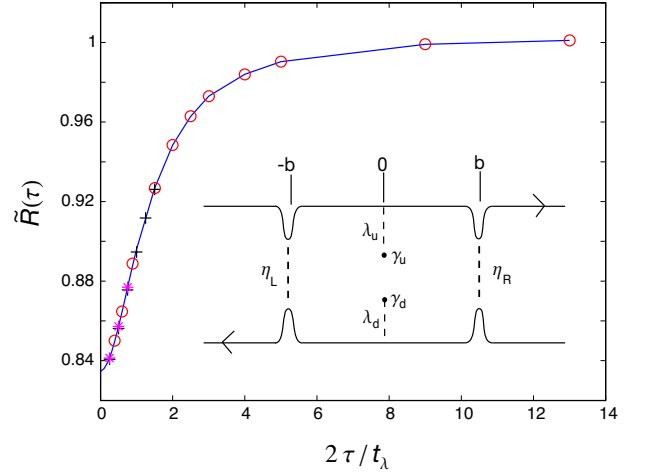}
\caption{ Reduction $\tilde{R}(\tau)$ of the imaginary time parity-parity correlation
function due to coupling between edge and localized modes.
Inset shows the interferometer with two
localized Majorana modes $\gamma_u$, $\gamma_d$ coupled to upper and lower edge, respectively. Data were numerically obtained for an interferometer of size $b=t_\lambda/4$ and total system size
 $L + b = 10.5 t_\lambda$, where $t_\lambda^{-1}$ is the characteristic
 decay rate of localized modes.
Data points are shown for $a/t_\lambda = 0.005$ (full line with open circles),
$a/t_\lambda = 0.0025$ (plus signs), and $a/t_\lambda = 0.00125$ (asterisks),
where $a$ denotes the lattice constant.   }
\label{reductionfactor.fig}
\end{figure}

In order to analyze the strong coupling limit with either $t_b$ or
$t_V$ of the order of $t_\lambda$, we numerically study a lattice
version of our model \cite{futurepaper}. We start with a tight-binding
Hamiltonian for one-dimensional complex fermions without the localized
modes: $H \ = \ - v_m a^{-1} \sum_j ( c_{j+1}^\dagger c_j \ + \
h.c. )$.
%
%
Here, $a$ is the lattice constant, and the operators obey the usual
anti-commutation relations $\{ c_i^\dagger, c_j\} =
\delta_{ij}$. We study the model at half filling with a Fermi wave
vector $k_F = {\pi / 2}$. The  fermions created by
$c_j^\dagger$ can be decomposed into two Majorana species
%
\begin{equation}
\gamma_{j} \ =\ e^{i {\pi \over 2} j} \ c_j \ + \ e^{- i {\pi \over
2} j}\ c_j^\dagger \ , \ \ \ \ \ \tilde{\gamma}_{j} \ = \ {1 \over
i} \big( e^{i {\pi \over 2} j} \ c_j \ - \ e^{- i {\pi \over 2} j}\
c_j^\dagger \big)  \, .  \nonumber
\end{equation}
%
We define continuum fermions, using $\gamma_j$ alone, via
%
\begin{eqnarray}
\Psi^{u,d}(j a) & = & {1 \over \sqrt{a}}   \sum_{j^\prime} (\pm 1)^{j^\prime} f(j - j^\prime) \gamma_{j^\prime}  \ \ ,
\label{lattice_continuum.eq}
\end{eqnarray}
%
where $f(j)$ is a Gaussian with a width large compared to the lattice
spacing and subject to the normalization $\sum_j f(j) = \sqrt{\pi}$.
Using this mapping one can now include the coupling to the localized
modes as in Eq.~(\ref{bulk-edge-tunneling}).

The parity operator for a set of lattice sites $\{j\}$ can be written
as the product over sites of operators $2 c_j^\dagger c_j -1 = i
\gamma_j \tilde{\gamma}_j$. The parity operator for the localized
modes has a similar form.  The expectation value of any such product,
at the same or different times, can be evaluated, using Wick's
theorem, as the Pfaffian of a matrix whose elements are the pair
correlation functions of operators on different sites, including the
localized modes where appropriate.  As the Hamiltonian is a sum $H =
H_\gamma + H_{\tilde{\gamma}}$, the $\tilde{\gamma}$ species
contributes a factor to the parity expectation value which is the same
whether the localized modes are present or not.  Thus we may ignore
the $\tilde{\gamma}$ modes in evaluating the reduction factor.  In
this way, the lattice form for the edge parity for a region
$[x_1,x_2]$ becomes
%
\begin{equation}
P_{edge}(x_1,x_2) \ = \ \prod_{x_1 \leq x_j \leq x_2} \sqrt{i}\;
\gamma_j \ \ . \label{parity-lattice}
\end{equation}

To calculate the reduction factor at time $t$ we will need to evaluate
expectation values of products like
%
$
\langle P_{edge}(-L, -b; t)  P_{edge}(-L, b; 0) i \gamma_u \gamma_d  \rangle \, ,
$
where $-L \ll - |t|$ is a point far to the left of the origin.  In the
absence of bulk-edge coupling, the pair correlation function for two
lattice points at equal times is
%
\begin{equation}
\langle i \gamma_j \gamma_k \rangle_0 \ = \ {1 \over \pi} {1\over j - k} \ \Big[
1 \ - \ e^{- i \pi (j - k)} \Big] \ \ .
\label{twoparticle.eq}
\end{equation}
%
The two terms in brackets arise from the right-moving upper edge and
left-moving lower edge, respectively.  With non-zero bulk-edge
coupling, the correlation functions $\langle i \gamma_j \gamma_k
\rangle$ and $\langle i \gamma_j \gamma_{u(d)}\rangle$ can be
calculated analytically in the continuum limit (i.e., for points not
too close to the origin) for the same and for different times.  A full
lattice calculation can be carried out numerically, but it is
time-consuming for large lattices.  We have found that the
short-distance errors introduced by using continuum correlation
functions only lead to an error in the Pfaffian by a factor
$c_\lambda$ that depends on $a/t_\lambda$ but is independent of the
interferometer size $b$ and the time difference $t$.  The numerical
results presented in Fig.~1 were obtained using the continuum
correlation functions and corrected by the factor $c_\lambda$ . The
numerical value of $c_\lambda$ can be obtained most easily by
considering equal time correlation functions.  As a check, we note
that results for different values of $a /t_\lambda$ obtained with this
numerical technique deviate less than 0.3 \% from each other, as shown
in Fig.~1.

Fig.~\ref{reductionfactor.fig} displays the {\em imaginary time}
reduction factor $\tilde{R}(\tau)$ for intermediate and strong
coupling, for an interferometer size $ b = t_\lambda/4$.
$\tilde{R}(\tau)$ is related to the real time reduction factor via the
analytic contiuation $R(t)=\tilde{R}(\tau \to i t +
\delta)$. The reduction factor $\tilde{R}(\tau)$ monotonically
increases with increasing time. At $\tau \approx t_\lambda$, there
is a crossover from parity reduction determined by the
interferometer size $b = t_\lambda/4$ to parity reduction
determined by the time. At large times $\tilde{R}(\tau)$ seems to
saturate at a value of one, implying that its analytic
continuation $R(t)$ saturates near one as well. Note that when
$R(t)=1$ the visibility of the interference is the same as it
would have been in the absence of the two bulk quasiparticles.
Similar results are expected if we have one strongly coupled
localized mode inside the interferometer path, and a second
localized mode, of arbitrary coupling, outside the interferometer.
We attribute the re-emergence of the
interference as the bulk-edge coupling gets strong to the
correlations that develop between the occupation of the fermionic
mode associated with the two quasi-particles and the occupation of
the region of the edge at a distance $v_mt_\lambda$ from the
coupling point. Each of these occupations strongly fluctuates due
to the coupling, but their fluctuations are strongly correlated.

In conclusion, we have found that when the coupling between the
Majorana mode associated with a localized $e*=e/4$ charged
quasiparticle and an adjacent edge is sufficiently strong, so that the
characteristic tunneling time $t_\lambda$ is short compared to the
time scale $t_V = \hbar e^* / V$ set by the voltage, it appears as if
the localized quasiparticle has become part of the edge.
Specifically, for an interference path enclosing the quasiparticle,
the interference visibility should have essentially the same strength
as if the quasiparticle were not there.  For weak coupling, the
time-averaged interference intensity is reduced, by a factor which is
$\propto (t_V/t_\lambda) \log^2 (t_\lambda /t_V)$ in the case where
there are two localized quasiparticles inside the loop, coupled
respectively to the two edges with similar strength.

{\em Acknowledgments:} We would like to thank B.J.~Overbosch and
X.-G.~Wen for fruitful discussions. This work was supported in part by
the Heisenberg program of DFG, by NSF grant DMR-0541988, the US-Israel
BSF, the Minerva foundation, and the Israel Science Foundation.


\begin{references}
\bibitem{endnote2} A different gauge choice would have given us the
parity of the number of electrons to the right of $x_0$.
\bibitem{endnote3} Since our reduction factor R(t) is a product of separate contributions from the
upper  and lower  edges, we can infer from one of these 
the result for a single localized mode.
\bibitem{futurepaper} More details will be given in a future
paper.
\end{references}
\end{document}